\documentstyle[epsf]{article}
\begin{document}
\renewcommand{\thefootnote}{\fnsymbol{footnote}}
\begin{center}
{\bf \Large
A POSSIBLE CORRELATION BETWEEN EGRET SOURCES
AND AN AIR-BORNE EXPERIMENT
}\\
\vskip 0.5cm
{Ryoji ENOMOTO\footnote{Internet address: enomoto@kekvax.kek.jp}
\\
\vskip 0.5cm
{\it National Laboratory for High Energy Physics, KEK \\
1-1 Oho, Tsukuba-city, Ibaraki 305, Japan}}

\end{center}
\vskip 1cm
\begin{abstract}

In 1989, an air-borne experiments (VEGA experiment) aiming at the
detection of a few 10 GeV $\gamma$-ray were carried out. In
these experiments, nine point-source candidates along the
Galactic plane were reported. In these candidates, the five of
five highest significance candidates positionally coincide with the
EGRET galactic plane sources.
\end{abstract}

\section{Introduction}

Recently, the EGRET(CGRO) developed a new era of the GeV-$\gamma$-ray
astrophysics. More than ten times of point sources were discovered by
this satellite \cite{bib1}.
On the other hand, in 1989, we carried out an air-borne experiment
for detecting a few ten GeV $\gamma$-rays \cite{bib2,bib3,bib4}.
In the first experiment,
there was an indication of the existence of the galactic plane sources.
We report a possible correlation between the EGRET sources and the
air-borne experiment which was first noticed by J. R. Mattox (GSFC).

\section{VEGA Experiment}

The experimental principle of the air-borne experiment (VEGA experiment)
are as follows. High momentum secondary electrons ($>$1 GeV) inside air
showers have almost the same directions with respect to the incident
$\gamma$-rays. The ambiguities of the direction are due to the multiple
scattering and the geomagnetic field, which are estimated to be
$\sim$0.5 degree. The detail can be found in Reference \cite{bib2}.

In the experiments, we used the flight between Narita (Japan) and
Sidney (Australia). The field of view was focused at the galactic
plane between $l$=-20 to 60 degrees. In the first experiment, we
have detected some fluctuation concentrated along the galactic
plane \cite{bib2}.

\section{Possible Correlation}

\begin{figure}
\epsfysize9cm
\epsfbox{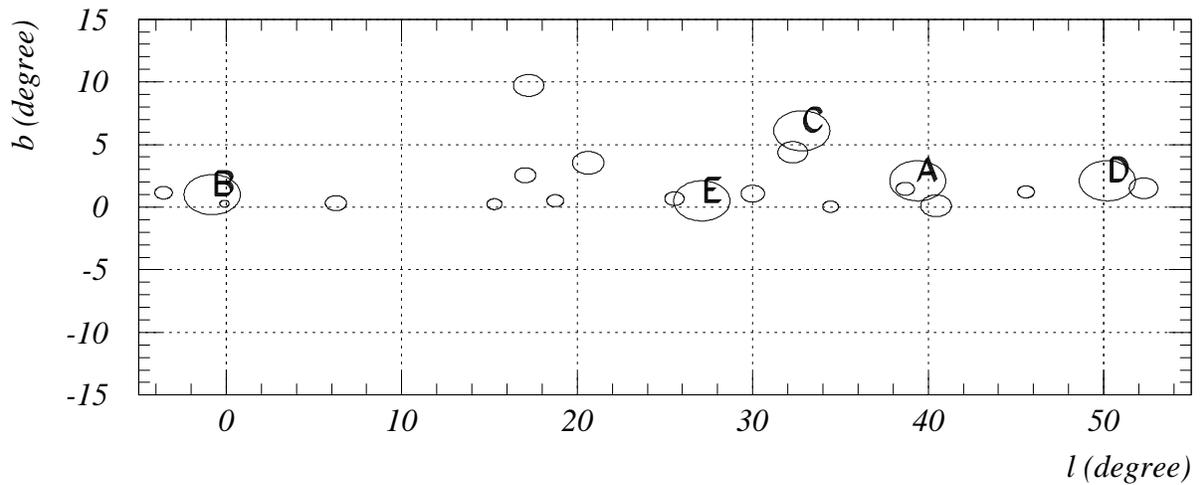}
\caption{Positions of galactic plane sources. The circles
represent the measurement errors. The circles marked by the alphabets
indicate the VEGA source candidates.}
\label{fcorr}
\end{figure}

The galactic plane sources of the EGRET are catalogued in Reference
\cite{bib1},
in Table 4a, and 4b. They are plotted in Figure \ref{fcorr}.
The typical error circles are 30 arcmin which are indicated with circles
in Figure \ref{fcorr}. The five highest significance peaks by the VEGA
experiment are also shown in Figure \ref{fcorr} tagged by A, B, C, D, and E.
The magnitudes of the error circles are 1.6 degree \cite{bib4}.
All of five positionally coincide with the EGRET sources which are listed
in Table \ref{tcorr}.
\begin{table}
\begin{center}
\begin{tabular}{ccc}
\hline
\hline
VEGA source-ID & EGRET source-ID & Comment \\
\hline
A & GRO J1857+05 & 1),2) \\
B & GRO J1732-31 & \\
C & GRO J1834+01 & 2) \\
D & GRO J1921+17 & 1) \\
E & GRO J1835-06 & 2) \\
\hline
\hline
\end{tabular}
\end{center}
\caption{Relations between the VEGA source candidates and the EGRET
sources. The comments are 1) extended or multiple sources, and 2)
transient source.}
\label{tcorr}
\end{table}
The three of five sources are transient and two are considered to be
extended or multiple sources by the EGRET observation.
The probability of the accidental coincidence are considered to be 2\%.

\section{Discussion}

In spite of a positional coincidence, there are two problems as follows.
\begin{enumerate}
\item The intensities of these sources by the VEGA experiment are
more than 10 times bigger than those obtained by the EGRET.
\item In the succeeding VEGA experiments, the existence
of these sources could not be confirmed \cite{bib4}.
\end{enumerate}
However, by checking carefully on the distributions of the electron arrival
directions in Reference \cite{bib4}, it is possible to say that there are
some fluctuations in these signal regions. Therefore if the positional
coincidence is a real effect, these sources must be strongly time dependent.
It is very interesting to observe these by different energy regions
such as X-ray and VHE-$\gamma$-ray.
\section{Summary}

A possible correlation between the EGRET sources and
the air-borne experiment
was found. The five of five highest significance source candidates
detected by
the VEGA experiment positionally coincide with the EGRET galactic
plane sources. Three of five were transient sources.
Efforts for establishment, such as X-ray and
air-\v Cerenkov observations, are important.

\section*{Acknowledgement}

This correlation was first noticed by J. R. Mattox (GSFC)
and I greatly appreciate it.
This work was supported by a
Grant-in-Aid in Scientific Research in Priority Areas from
the Japan Ministry of Education, Science, and Culture.

\end{document}